\documentclass{ptapap}
\usepackage{amsmath}
\usepackage{amssymb}

\author{Wojciech A. Hellwing}[CFT,PAS]
\affil[CFT]{Center for Theoretical Physics, Polish Academy of Sciences, Al. Lotników 32/46, 02-668 Warsaw, Poland}
\affil[PAS]{Polish Astronomical Society, Bartycka 18, 00--716 Warsaw, Poland}

\title{The skewness of $z=0.5$ redshift-space galaxy distribution in Modified Gravity}

\usepackage{times,graphicx,amsmath,amsfonts,amssymb,aas_macros,epstopdf,ulem}
\usepackage{epsfig}
\usepackage[usenames,dvipsnames]{xcolor}
\usepackage{url}
\usepackage{subfigure}
\usepackage[colorlinks=true, urlcolor=black, citecolor=blue]{hyperref}
\usepackage[utf8]{inputenc}

\def\ie{{\frenchspacing\it i.e.}~}
\def\eg{{\frenchspacing\it e.g.}~}

\def\be{\begin{equation}}
\def\ee{\end{equation}}
\def\ba{\begin{eqnarray}}
\def\ea{\end{eqnarray}} 
\def\Msun{h^{-1}{\rm M}_{\odot}}
\def\hmpc{h^{-1}\,{\rm Mpc}}

\def\dd{\textrm{d}}

\newcommand{\av}[1]{\langle{#1}\rangle}
\newcommand{\xa}[1]{\overline{\xi}_{#1}}
\def\frac#1#2{{\textstyle{#1\over #2}}}

\def\simlt{\stackrel{<}{{}_\sim}}

\def\lcdm{\ensuremath{\Lambda}CDM{~}}

\begin{document}

\maketitle
\begin{abstract}
We study the reduced skewness, $S_{3,g}\equiv\xa{3,g}/\xa{2,g}^2$ of galaxy distribution at $z=0.5$ in two families of modfied gravity models:
the Hu-Sawicki $f(R)$-gravity and normal-branch of Dvali-Gabadadze-Porrati (nDGP) models. We use a set of mock
galaxy catalogues specifally designed to match CMASS spectroscopic galaxy sample.
For the first time we investigate the third reduced moment of such galaxy distributions
in the redshift space. Our analysis confirms that the signal previously indicated only for dark matter halo catalogues
persists also in realistic mock galaxy samples. 
This result offers a possibility to extract a potential modified gravity signal in $S_3$ from spectroscopic galaxy data without
a need for a very precise and self-consistent RSD models constructed for each and every modified gravity scenario separately.
We show that the relative deviations  from {\ensuremath{\Lambda}CDM{~}}
$S_{3,g}$ of various modified gravity models can vary from $7$ down to $\sim 2-3\%$ effects. 
Albeit, the effect looks small, we show that for considered models it can foster a $2-3\sigma$ falsification.
Finally we argue that galaxy sample of a significantly higher number density should provide even stronger constraints by
limiting shot-noise effects affecting the $S_{3,g}$ estimates at small comoving separations.
\end{abstract}

\section{Introduction}
\label{sec:intro}
The standard model of cosmology, namely the Lambda-Cold-Dark-Matter model (\lcdm), where the observed late-time accelerated expansion 
taken together with the core-assumption
of the General Relativity (GR) being the correct theory of gravitation on all scales and for all epochs, implicate that more than two-thirds
of the current cosmic energy-density budget is made-up by a mysterious Dark Energy.
The assumption that GR provides adequate physical description of the Universe and holds over 27 orders of magnitudes in scales (from centimeters to gigaparsecs)
is a very strong one. The scientific method requires from us to put such assumptions under rigorous and constant scrutiny.

A contemporary approach to this problem consists of two complementary avenues: (i) designing and conveying various observational tests of GR on cosmological
and intergalactic scales; and (ii) to explore theoretical freedom to modify GR. The former is manifested in one of the key goals set for XXI-century
extragalactic astronomy.  Many ongoing and planned for the near future grand-design observational campaign and space missions, such as Euclid satellite, 
the Dark Energy Spectroscopic Instrument (DESI) survey or Large Synoptic Survey Telescope (LSST) have conveying test of GR and Dark Energy paradigm 
as their primary science cases [\cite{Aghamousa:2016zmz,LSST}. 
The latter avenue, have lead to a discovery of many self-consistent, pathology-free (such as \ie{} ghost-states)
models that go beyond-GR. These have been commonly dubbed as Modfied Gravity (MG) theories \cite{Clifton2012}. A unique property of the most viable MG models, that sets them
apart from GR is that they feature propagation of extra degrees of freedom (d.o.f.). When coupled to matter fields these would manifest as a universal 
fifth-force acting only in intergalactic and cosmological, but not on smaller scales. The latter is thanks to a non-linear screening mechanisms that MG models
employ to suppress the fifth-force on small-scales and in the strong-field regimes, where we currently have precise tests of the gravity theory 
[\cite{Chiba2007,Abbott2016, Monitor:2017mdv,GBM:2017lvd}].
To this end interesting and viable MG theories contains new physics and predicts differences in growth rate and distribution of large-scale structures, compared
to the vanilla \lcdm model with GR. It is this feature that is exploited, when searching and designing observational test of GR and beyond-GR model.

In this paper, we consider the skewness of the galaxy Count-in-Cells distribution as one of large-scale structure tests of the theory of gravity.
For the first time we also compare the real and redshift-space variance and the skewness of mock galaxy catalogues designed to model clustering of
galaxies at redshift $z=0.5$, as will be measured by large and deep surveys such as DESI. 

\section{Modified Gravity}
\label{sec:mg}
Alternatives to \lcdm are numerous (see e.g. review \cite{Clifton2012,Joyce2015,Koyama2016}), but numerous also are the problems they have to struggle with:
some of them are plagued with theoretical instabilities and all of them have to face observational constraints which often require 
fine-tuning of model parameters.
In this work we consider screened modified gravity models, where the extra fifth force is screened in high-density (or high-potential) regions. 
Specifically, we study two classes of such screened modified gravity models: the $f(R)$-gravity and nDGP\footnote{nDGP stands for the normal-branch 
Dvali-Gabadadze-Porrati model.} braneworld gravity. These two models constitute a very good test suite for a wider class of modified gravity theories.
This is because most of the viable MG models can be divided into two general categories, depending on the physical mechanism of the fifth-force screening
they invoke. The screening can be either environmentally dependent or object mass dependent. The former mechanism is responding to the local value of 
the gravitational potential, in the latter (also called the Vainshtein mechanism) the effectiveness of the screening is usually moderated by the local 
curvature of a given region of space. 

\subsection{$f(R)$ gravity}
\label{subsec:f_of_R}
The $f(R)$ gravity is an extension of GR that has been extensively studied in the literature in the past few years (see \cite{Sotiriou2010}, for detailed reviews). 
The theory is obtained by substituting the Ricci scalar $R$ in the Einstein-Hilbert action with an algebraic function $f(R)$. Here, the model can
be tuned to give an accelerated expansion produced by this extra term replacing cosmological constant ($\Lambda$) in the action integral and
consistent with the \lcdm{} expansion history (\eg{} \cite{HuSawicki}). 
The resulting  modified theory of gravity is characterized by highly non-linear equations of motion for the scalar field and environmentally dependent fifth-force 
screening is obtained via the so-called {\it chameleon mechanism}. Hitherto the screening is achieved by effectively making the scalar field very massive
in the locally dense regions, making the field’s Compton length very small and thus effectively suppressing the fifth force. In contrast, in 
the low density regions the scalar field retains a low mass and its gradient can produce significant fifth force effectively enhancing the local 
Newtonian gravity. In this class of models, the degree of potential deviations from the LCDM dynamics is controlled by the $|f_{R_0}|$ parameter,
which is a normalized scalar-filed amplitude at present time. We consider three cases of the Hu-Sawicki $f(R)$ model that varies by an order of magnitude
between each other in that parameter. Specifically, we choose $|f_{R_0}|=10^{-6},10^{-5}$ and $10^{-4}$ and dub them respectively as F6, F5 and F4.

\subsection{Braneworld DGP model}
\label{subsec:nDGP}
The nDGP gravity, the second model we study, belongs to the so-called braneworld gravity models. This picture, inspired by string theory, 
fosters the observed Universe as a four-dimensional brane embedded in a higher-dimensional bulk space-time. Matter fields are confined to 
the brane, while gravity can propagate in the whole N-dimensional bulk space-time. In this class of scenarios the accelerated expansion is 
realised via higher-dimensional effects rather then by dark energy. A classical example of such a braneworld model is the Dvali-Gabadadze-Porrati 
(DGP) model [\cite{DGB2000,Koyama2007}]. A natural extension of the DGP model consists of a universe with higher-dimensional bulk space-time. As a test case 
scenario we choose to study the so-called normal branch DGP model (nDGP)\cite{Sahni2003}. This extension of the DGP scenario still requires some amount 
of GR-like dark energy, hence is a bit less appealing theoretically. However, it is still consistent with current observations and exhibits 
the second kind of the screening -- the Vainshtein mechanism. Here the non-linear self-interactions of the additional scalar degree of freedom 
are able to shield the fifth-force on small scales. As the Vainshtein screening only depends on the mass of an object and the distance from it, 
the fifth-force produced in a gravity models implementing it will have, in general, a different magnitude and scale of action compared to 
models with chameleon screening.
In the brane-world class of models a parameter that determines the behavior of the model and the strength and scale of potential departures 
from GR-based predictions is the so-called cross-over scale, $r_c$. Which, when expressed in the units of the present-day Hubble constant $H_0$
corresponds to a scale at which the effects of the 5-th dimension space-time for gravity starts to be significant. In this work we study
two variants of the nDGP with $r_cH_0=1$ and $5$ Gpc/h. We mark those two version as N1 and N5 respectively.

\section{Simulations and mock catalogues}
\label{sec:sims}
For our analysis we will use mock galaxy catalogues build-on the input dark matter halo catalogues from 
{\sc ELEPHANT} (Extended LEnsing PHysics with ANalytical ray Tracing) suite \cite{Elephant}. This is a set of dedicated
N-body dark matter-only simulations. While such a gross simplification will not admit for modeling of any intrinsic galaxy properties,
such as their colours, morphology, metallicity or luminosity, it is sufficient for modeling realistic spatial distribution of galaxy
samples.

For all our  simulations we choose the same background cosmology that of \lcdm{} model described by a best fit WMAP9 cosmology \citep{WMAP9}.
These are the following: $\Omega_M=0.281$ (present fractional matter density), $\Omega_\Lambda=0.719$ (present dark energy fractional density),
$h=0.697$ (present-day Hubble constant $H_0/(100$km s$^{-1}$Mpc$^{-1}$),
$n_s=0.971$ (primordial power spectrum index), $\sigma_8=0.82$ (present day linear power spectrum amplitude normalization).
For each of our six models (\ie{} GR, F4, F5, F6, N1 and N5) we run 5 independent realisations of initial conditions, which are shared
among models. We will average all results for any given model over those 5 realisations to reduce the impact of the cosmic variance.
Finally, all the runs were conducted with a use of $1024^3$ pseudo-particles to sample dark matter phase-space in a periodic
cubic box of $1024\hmpc$ on a side, resulting in the mass resolution of $m_p=7.78\times 10^{10}\Msun$.

To identify gravitationally bound dark matter structures -- dark matter haloes, we use the publicly-available {\sc rockstar} 
halo finder\footnote{https://bitbucket.org/gfcstanford/rockstar}\citep{Behroozi2013}. These, are primary sites for galaxy formation. 
To map the halo catalogs to a corresponding galaxy distribution, we resort to the Halo Occupation Distribution (HOD) method [\citep{Berlind:2002rn,Zheng:2004id}].
Here, the main assumption is that the probability for a halo to host a certain number of galaxies can be estimated via a simple functional dependence 
on the mass of the host halo. Specifically, we implement the HOD variation of \cite{Zheng2007}, where the mean number of 
central galaxies, $\left< N_{{\rm cen}} (M) \right>$, and the mean number of satellite galaxies, $\left< N_{{\rm sat}} (M) \right>$, 
in a halo of mass $M$, are chosen so that
to maximise the fit to desired number density of galaxies, $n(z)$, and the projected galaxy two-point correlation functions (2PCFs), $w_p(r_p)$. For a specific galaxy catalogue to model we take the characteristics of $z=0.5$ the CMASS data, as described by their HOD model by
\cite{Manera2013}. This choice is dictated by the fact that our intention is to study the real and redshift-space distribution of galaxies as can be collected
from the current and forthcoming large sky surveys. 

\section{Skewness and the hierarchical clustering}
\label{sec:skew}
In what follows we will study the galaxy distributions 1-point central moments, with a specific emphasis of the third moment, the skewness. 
We start by denoting that the cosmic smooth density field can be described by the statistics for one random variable. The density contrast
$\delta_m(\vec{x})=\rho(\vec{x})/\rho_b-1$. Where $\rho_b$ is the mean background density and $\rho(\vec{x})$ is the local density at a co-moving
location $\vec{x}$. By smoothing this density field with a spherical Top-Hat filter, $W_R(\vec{x})$, we get a density contrast field evaluated at a specific
smoothing scale $R$.

In cosmologies with nearly scale-invariant primordial power spectrum with the non-relativistic dark matter, the matter and galaxies distributions
forms a hierarchy of connected moments \cite{Fry1984a,Fry1984b}
\begin{equation}
 \label{eqn:connected-moments}
 \bar{\xi}_=\av{\delta^n}_c= S_j\xa{2}^{j-1}\,,
\end{equation}
where $\xa{2}=\av{\delta^2}_c$ (thus the variance) and the bar denotes that the averaging is over the volume (\ie over spheres). Here, $S_j$'s 
are the so-called hierarchical amplitudes or reduced moments (of the $j-$th order)  and we will be specifically interested in $S_3$ -- the reduced skewness. 
\cite{FG1993} have shown that the matter and galaxies connected moments are related by:
\begin{equation}
 \label{eqn:biased_moments}
 \xa{2,g}=b^2\xa{2} + \mathcal{O}(\xa{2}^2)\,\quad \bar{\xi}_{3,g}= b^3\xa{2}^2(S_3+3c_2) + \mathcal{O}(\xa{2}^3)\,,
\end{equation}
where $b\equiv b_1$ is the usual linear bias, $b_2$ is its second order term in Taylor expansion and we have defined $c_2\equiv b_2/b$.

\cite{Juszkiewicz1993} have shown that the skewness of a smoothed matter density field is a weak function of the smoothing
scale, $R$, with a dependence on the effective slope of the power-spectrum (\ie{} variance) at that smoothing scale:
\begin{equation}
 \label{eqn:s3_smoothed}
  S_3={34\over 7} + \gamma_1\,,\qquad \textrm{where}\,\, \gamma_1\equiv{\dd\textrm{log}\sigma^2(R)\over\dd\textrm{log} R}\,.
\end{equation}
Since the $\gamma_1$ is a function of the matter variance shape, the changes here induced by the 5-th force of MG are reflected in
modified values of the skewness in MG scenarios. This has been shown to be a sensitive probe of the MG (\cite{Hellwing_npoint, Hellwing2013}).
Now, since the galaxy distribution skewness is a function of $S_3$ and the bias factor, ref. Eqn.(\ref{eqn:biased_moments}):
\begin{equation}
 S_{3,g}=b^{-1}(S_3+3c_2)\,,
\end{equation}
we can expect that this statistic should also be different for beyond-GR models. Unless, there would be a conspiracy between $b$
and $c_2$ values to exactly cancel-out any deviations form GR. Which for  the case of dark matter haloes has been shown to not be the case (\cite{Hellwing2017}).
\begin{figure*}
  \centering
  \begin{minipage}{0.48\textwidth}
    \includegraphics[width=\textwidth]{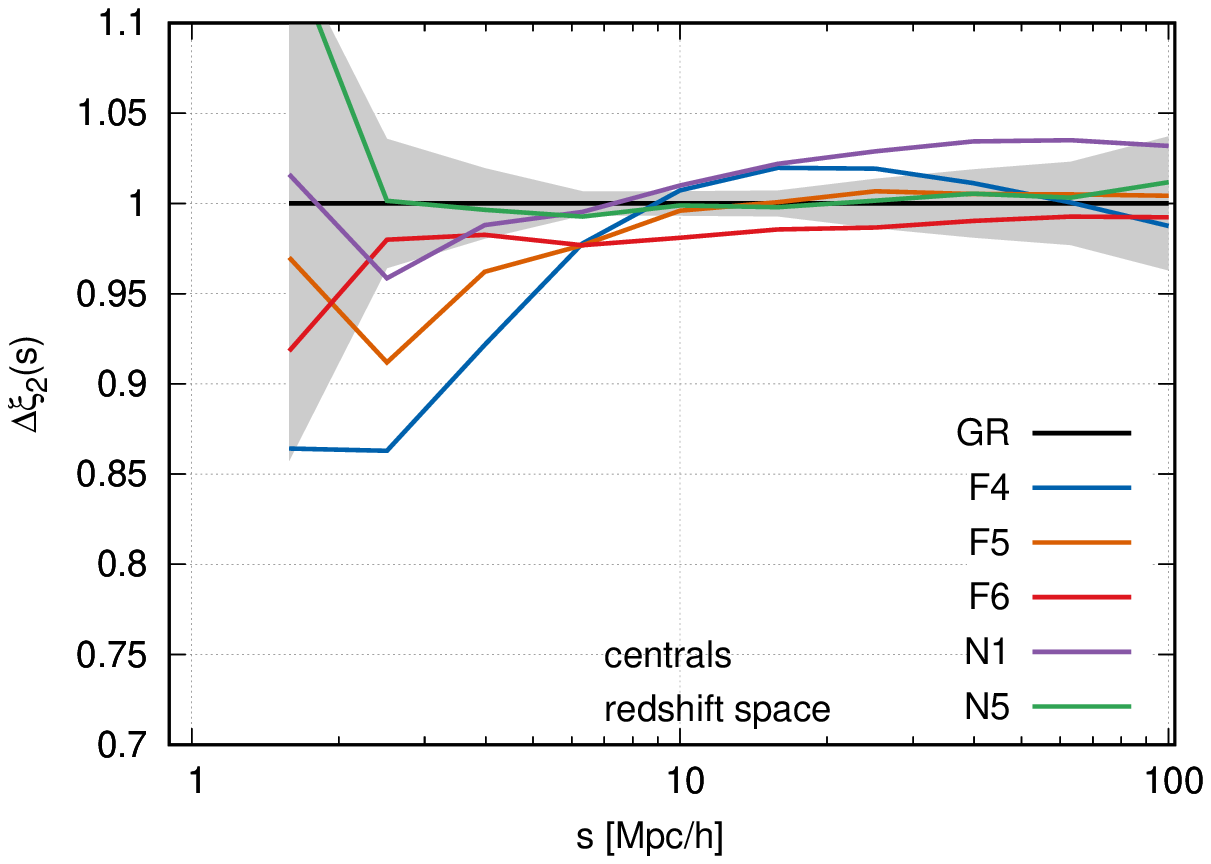}
    \caption{The deviation of redshift-space variance $\xa{2,g}(s)$ w.r.t. GR case taken for central galaxies distributions.
    The shadow region marks a typical ensemble average error of the ratio.}
    \label{fig:xi2_ce_RSD}
  \end{minipage}
  \quad
  \begin{minipage}{0.48\textwidth}
    \includegraphics[width=\textwidth]{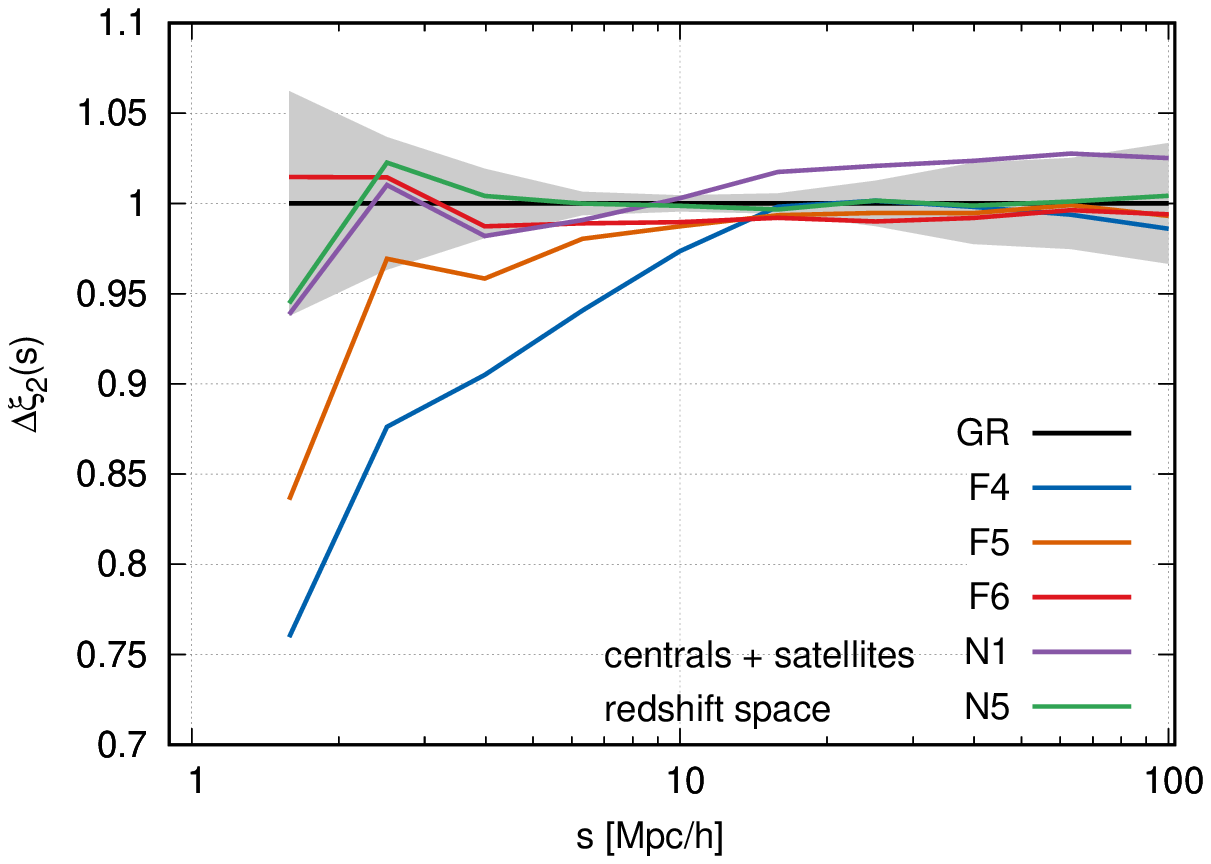}
    \caption{Same as the left figure but for centrals + satellites HOD catalogues.The shadow region marks a typical
    ensemble average error of the ratio.}
    \label{fig:xi2_all_RSD}
  \end{minipage}
\end{figure*}

Finally, we should discuss the potential effects that will be induced by the measurements being made in the redshift-space, where 
the line-of-sight component of the peculiar velocity perturbs the inferred distance to an object.
Here we resort to a distant-observer approximation, where all line-of-sight from observes to galaxies in simulations are assumed to be
parallel. This gives for sources at cosmological redshift of $z\simeq 0.5$ a typical displacement due to peculiar velocity 
of the order of $\sim 1.18 (v_{||}/100$km/s$)$Mpc/h. This is very small compared to 
the radial co-moving distance at that redshift $r(z\simeq 0.5)\simeq 1.32$Gpc/h, but is enough to induce significant effects on matter
and galaxy clustering statistics [\cite{FoG:Jackson,Kaiser:1987}]. However, \cite{Hivon1995} have shown that to the first order 
the connected moments $\xa{3}$ and $\xa{2}$ are affected by the redshift-space projections by nearly the same factor. Thus, the overall effect
for the skewness largely cancels-out. For example for the smoothed matter density in the flat \lcdm{} we have $|S_{3,z}-S_3|/S_3\simlt 0.2$.

\section{Results}
\label{sec:results}
{\it Moments estimation.} We will be analyzing the moments of the galaxy real and redshift space distributions. For a set of discrete tracers, such as galaxies,
the moments can be readily estimated using count-in-cells  (CIC) method [\cite{GaztanagaAPM94,cic_ana,BCGS_book}]. The counts 
define a discrete sample of the density distribution. 

Using this procedure we estimate $\xa{2,g}$, $\xa{3,g}$ and $S_{3,g}$ for our galaxy samples. For each of our simulations 
we use $5\times10^{5}$ randomly placed in the volume to sample the underlying $k$-moments at a given radius scale bin $R$.

\label{subsec:moments_rsd}
{\it Moments in redshift-space} Now, we will study the variance and the skewness of the galaxy distribution in the redshift-space.
Given the fact that nearly all large galaxy catalogues with a 3D galaxy positions are of a spectroscopic nature, it is crucial to
test the predictions of MG models for galaxy distribution in redshift-space. To obtain redshift space data, we take our real-space HOD 
and apply to them the RSD transformation taking independently x,y and z axes as  the line-of-sight axis. 
For our chosen redshift sample of $z=0.5$ it is still a reasonable approximation for our purposes.
\begin{figure*}
  \centering
  \begin{minipage}{0.48\textwidth}
    \includegraphics[width=\textwidth]{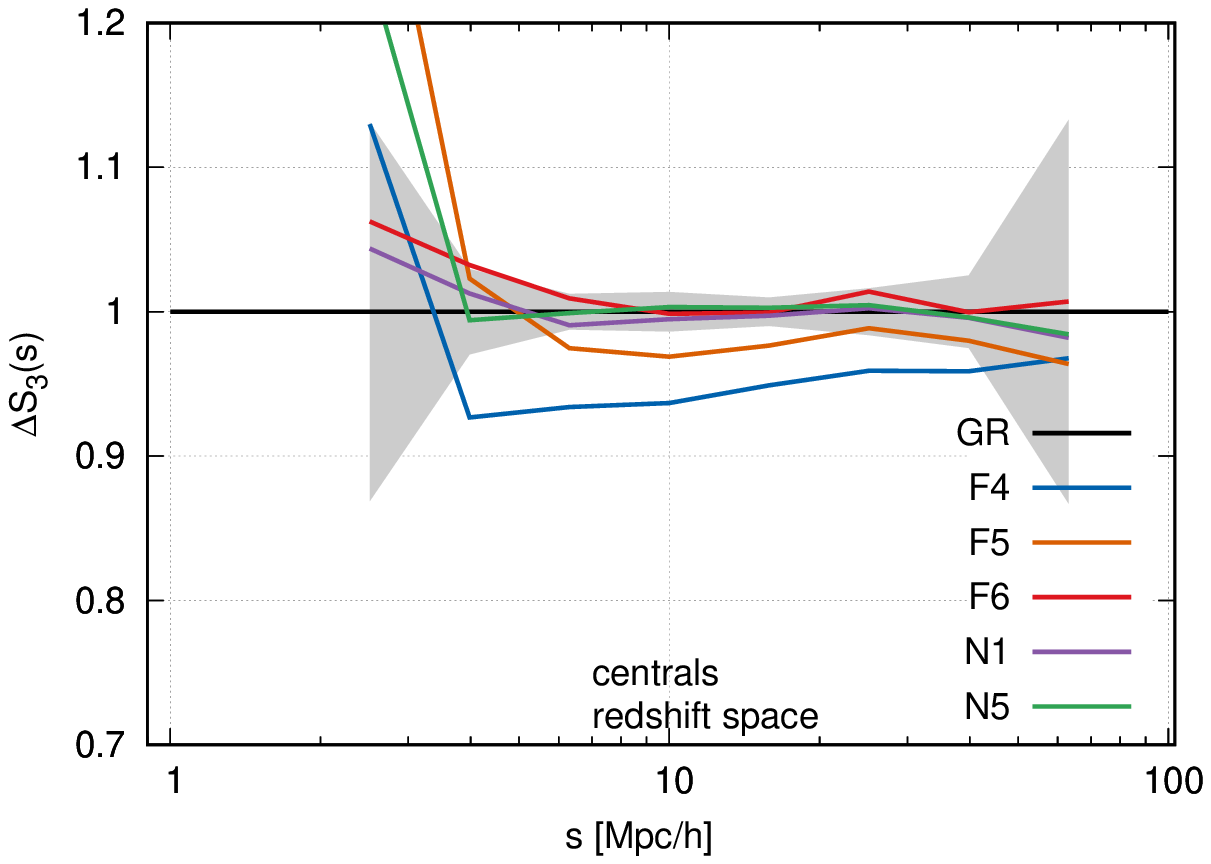}
    \caption{The deviation of redshift-space skewness $S_{3,g}$ w.r.t. GR case taken for central galaxies distributions. The shadow region marks a typical
    ensemble average error of the ratio.}
    \label{fig:s3_ce_RSD}
  \end{minipage}
  \quad
  \begin{minipage}{0.48\textwidth}
    \includegraphics[width=\textwidth]{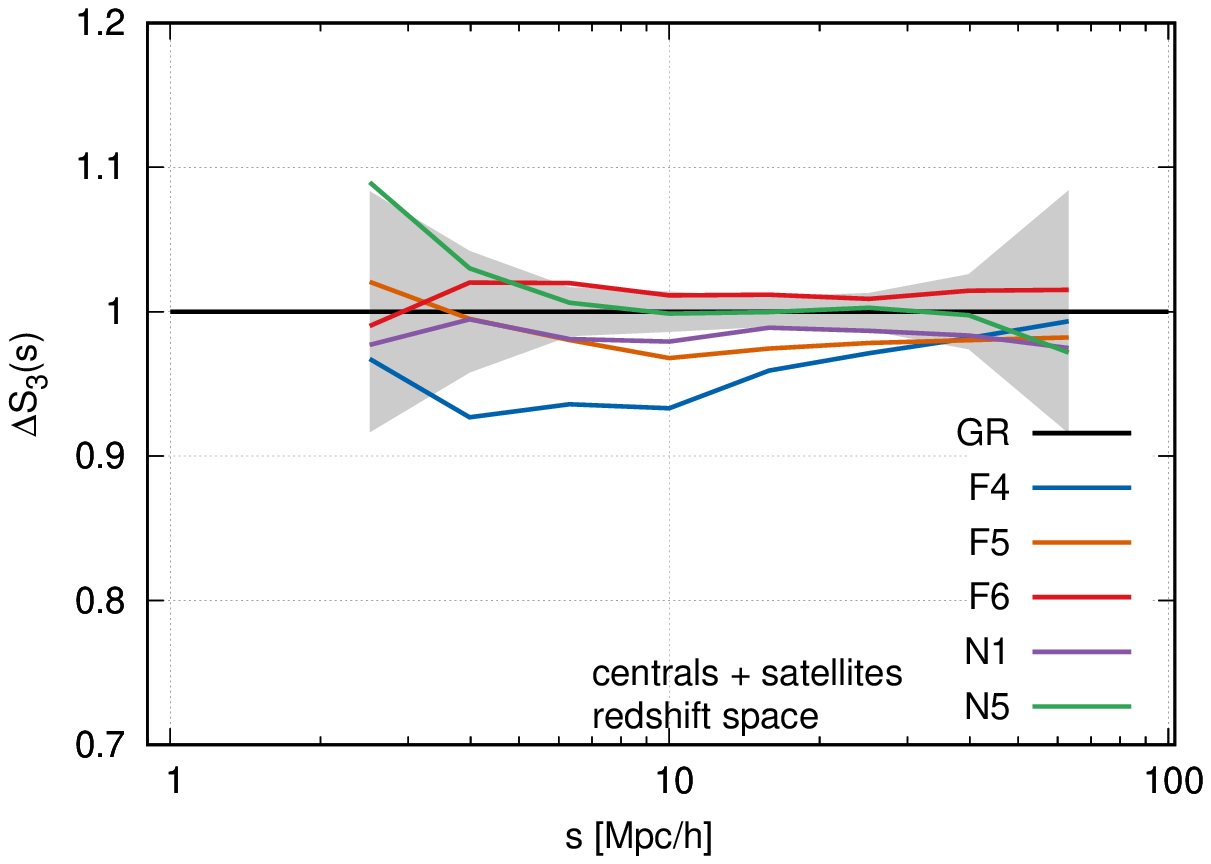}
    \caption{Same as the left figure but for centrals + satellites HOD catalogues.The shadow region marks a typical
    ensemble average error of the ratio.}
    \label{fig:s3_all_RSD}
  \end{minipage}
\end{figure*}

We start by investigating the variance of both types of galaxy catalogues. These are shown in Fig.~\ref{fig:xi2_ce_RSD} for the centrals
and in Fig.~\ref{fig:xi2_all_RSD} for the centrals+satellites respectively. Clearly we denote that the added effect of enhanced peculiar velocities impacts
even the variances of the resulting MG distributions. For the centrals only sample F4 and F5 shows the biggest differences at small radii $\simlt 10\hmpc$,
reaching a relative difference of the order of $\sim15$ and $\sim10$ percents at the smallest considered radius. This is a signal of $2-3\sigma$ level.
The N1 model is showing the nearly flat enhancement of $3-4\%$ at large scales $R\geq 10\hmpc$. The rest of the models
exhibit a small scatter around the GR-based values. 

Finally, we present the main point of our interest in this paper: the skewness of the redshift-space galaxy distribution. Traditionally,
in Fig.~\ref{fig:s3_ce_RSD} we plot the results for the centrals-only sample and in Fig.~\ref{fig:s3_all_RSD} for the full sample including
satellites.
We have compared these results to the corresponding relative devations we measured in the real-space (figures not shown here) and found out
that the overall magnitude of particular MG models departures from GR are comparable between both spaces.
This is in a clear agreement with the results of \cite{Hivon1995} obtained for the smoothed matter density skewness
in the redshift-space, where they shown that the overall effect of the redshift-space distortion is weak for the $S_{3}$. Our results indicate,
that the galaxy bias factors: $b$ and $c_2$ are not significantly affected by the redshift-space transformation and 
the corresponding redshift space reduced cumulants show a similar degree of deviation from the GR case. 
Despite the fact, that the overall relative differences predicted for various MG models are relatively small and contained to within $\simlt 7\%$,
they comprise a statistically significant signal for F4, F5  and N1 models at varying scales. Even F6, which is at the edge of the
1-sigma scatter band could, be potentially measured with a bigger volume sample.

\section{Summary}
\label{sec:summary}

We have investigated the variance and the skewness of $z=0.5$ mock galaxy distribution in redshift space projection. For that purpose
we employed a set of mock galaxy catalogues.
Our data can be regarded as a good proxy for a Luminous Red Galaxy (LRG) samples
at $z=0.5$ to be made available thanks to forthcoming space missions as Euclid and ground-base campaign like the DESI survey.

For the first time, we took a closer look at the differences of the reduced galaxy skewness $S_{3,g}$ between a few popular modified gravity models
and the fiducial GR-case. Thus our results here can be regarded as a first realistic forecast of the accuracy of such measurements to become
available with the incoming astronomical data.

Our analysis clearly demonstrates that there is a significant signal that screened modified gravity scenarios such as
the $f(R)$ and nDGP gravity models imprint in the skewness of the galaxy distribution. Previously, such a signature was only ascertained for
the distribution of simulated dark matter haloes [\cite{Hellwing2017}]. 
Even more importantly, the results we have obtained for the redshift-space reduced skewness,
$S_{3,g}(s)$ indicate that indeed the distortions caused by the spectroscopic distance projections affect to a similar degree both the variance
and the third connected moment $\xa{g,3}(s)$. Such observation is supported by the fact that similar modified gravity models features a comparable
degree of deviations from the \lcdm{} case in $S_{g,3}(s)$ as in $S_{g,3}(R)$. This observation opens a tantalizing possibility to extract
a potential MG signal form the reduced skewness of large spectroscopic galaxy samples. If such a measurement could be made successfully
(\ie{} avoiding any potential killer systematics) models such as F5, F4 and N1 could be falisfied wit ha modest $\sim2-3\sigma$ significant level.
In addition, if one could foster a galaxy sample with a significantly higher $n(z)$, thus alleviating the shot-noise errors at small separations,
an even tighter constraints could be obtained and also for a milder MG models such as F6 and N5.

\acknowledgements{
We are very grateful to Baojiu Li for providing us with the mock galaxy catalogues used in this work.
The Author is supported by National Science Center, Poland, under grant no 2018/30/E/ST9/00698.
This project has also benefited from numerical computations performed at 
the Interdisciplinary Centre for Mathematical and Computational Modelling (ICM) University of Warsaw 
under grants no GA67-17 and GA65-30.
}
\bibliographystyle{ptapap}
\bibliography{skewness_mg_pta_short}

\end{document}